\documentclass[a4paper,10pt]{article}
\usepackage{cite}
\usepackage{epsfig,amsmath,amssymb,verbatim,mathrsfs,array,layout,textcomp,amssymb,latexsym}
\usepackage{subfigure}
\newenvironment{Eqnarray}{\arraycolsep 0.14em\begin{eqnarray}}{\end{eqnarray}}
\def\beqa{\begin{Eqnarray}}
\def\eeqa{\end{Eqnarray}}
\def\beq{\begin{Eqnarray}}
\def\eeq{\end{Eqnarray}}
\newcommand{\no}{\nonumber}
\newcommand{\bv}{\left(\begin{array}{c}}
\newcommand{\ev}{\end{array}\right)}
\newcommand{\bmtwo}{\left(\begin{array}{cc}}
\newcommand{\bmthree}{\left(\begin{array}{ccc}}
\newcommand{\emn}{\end{array}\right)}
\newcommand{\bmtwoc}{\left\{\begin{array}{cc}}
\newcommand{\bmthreec}{\left\{\begin{array}{ccc}}
\newcommand{\emnc}{\end{array}\right\}}

\newcommand{\lag}{\mathcal{L}}

\newcommand{\abs}[1]{\left\lvert#1\right\rvert}
\newcommand{\avg}[1]{\left\langle {#1} \right\rangle}

\begin{document}
\begin{titlepage}

\vskip1.5cm
\begin{center}
{\Large \bf What if $\lambda_{hhh}\neq 3m_h^2/v$?}
\end{center}
\vskip0.2cm

\begin{center}
Aielet Efrati and Yosef Nir \\
\end{center}
\vskip 8pt
\begin{center}
{\it Department of Particle Physics and Astrophysics \\
Weizmann Institute of Science, Rehovot 76100, Israel} \vspace*{0.3cm}
\vskip 8pt
{\tt  aielet.efrati,yosef.nir@weizmann.ac.il}
\end{center}
\vskip 8pt

\begin{abstract}
A measurement of the Higgs trilinear self coupling $\lambda_{hhh}$ will test the Standard Model Higgs potential. But can it reveal information that cannot be learned otherwise? By analyzing several simple extensions of the Standard Model scalar sector we show that this measurement might give a first hint for New Physics modifying the electroweak symmetry breaking. Combining the measurements of $\lambda_{hhh}$ and $\lambda_{hVV}$ ($V=W,Z$) is particularly powerful in distinguishing between various models of New Physics and in providing unique information on these models.
\end{abstract}

\end{titlepage}

\section{Introduction}\label{sec:intro}
The recent discovery at the Large Hadron Collider (LHC) of a new scalar $h$~\cite{Aad:2012tfa,Chatrchyan:2012ufa} that couples to pairs of weak gauge bosons is an important triumph for the BEH mechanism of electroweak symmetry breaking \cite{brout,higgs}. The data analyzed so far by the ATLAS and CMS experiments suggest that the properties of this particle are consistent, within present experimental accuracy, with those of the Higgs boson of the Standard Model (SM). Yet, many well motivated New Physics (NP) scenarios predict the existence of additional new particles that mix with, or couple to the Higgs boson, leading to deviations of the Higgs couplings from the SM predictions. So far, experiments have given no direct hints for such particles, and it might be that the new particles are too heavy or too weakly coupled to be directly discovered even in the upcoming runs of the LHC. In this case, precise measurements of SM processes will be the only way to see their imprints. These two approaches, direct searches and precise measurements, are then complementary.

The SM Higgs potential is a two parameter model. One of them is the Higgs Vacuum Expectation Value (VEV), determined by the Fermi constant which was measured over eighty years ago, $v=(\sqrt{2}G_F)^{-1/2}\simeq246$ GeV. The other is the Higgs mass, measured by the CMS and ATLAS collaborations over the last year, $m_h=125.9\pm0.4$~GeV~\cite{Beringer:1900zz}. The measurement of $m_h$ exhausts the unknown parameters of the SM Higgs potential. In particular, within the SM, the trilinear Higgs self-coupling, $\lambda_{hhh}$, is not an independent parameter:
\beqa\label{eq:SMhhh}
\lambda_{hhh}^{\rm SM}=3m_h^2/v\,.
\eeqa
A future measurement of $\lambda_{hhh}$ will then {\it test} the SM Higgs potential. But can it reveal {\it new} information that cannot be learned otherwise? What can be learned if we find $\lambda_{hhh}\neq3m_h^2/v$? These are the questions we aim to address in this work.

Since we are interested in the case that no direct discovery of relevant particles is achieved, we examine various NP models in the decoupling limit. We analyze the relation between the Higgs self coupling and its couplings to the weak gauge bosons $V=W,Z$ and to the charged fermions $f=t,b,\tau$ at tree level:
\beq\label{eq:lhxx}
{\cal L}_{hXX}=
-\frac{1}{6}\lambda_{hhh}hhh+\lambda_{hVV}hW^+_\mu W^{-\mu}
+\frac12\lambda_{hVV}hZ_\mu Z^\mu-\lambda_{hff}h\bar ff.
\eeq
In all the models that we analyze, the NP respects custodial symmetry, and there is no need to distinguish between $\lambda_{hWW}$ and $\lambda_{hZZ}$. In all the models that we study, modifications to the Yukawa couplings are flavor-universal, and so there is no need to distinguish between the different generations in each sector.

Within the SM we have
\beqa\label{eq:SM}
\lambda_{hVV}^{\rm SM}&=&2m_V^2/v\,,\no\\
\lambda_{hff}^{\rm SM}&=&m_f/v\,.
\eeqa
We define:
\beqa\label{eq:defdel}
\delta\lambda_{hhh}&=&\frac{\lambda_{hhh}}{\lambda_{hhh}^{\rm SM}}-1\,,\no\\
\delta\lambda_{hVV}&=&\frac{\lambda_{hVV}}{\lambda_{hVV}^{\rm SM}}-1\,,\no\\
\delta\lambda_{hff}&=&\frac{\lambda_{hff}}{\lambda_{hff}^{\rm SM}}-1\,.
\eeqa
Our main focus will be on the lessons that can be learned from $\delta\lambda_{hhh}/\delta\lambda_{hVV}$.

A comment is in order regarding loop corrections to Eqs.~\eqref{eq:SMhhh} and~\eqref{eq:SM}. Within the SM, the relation between $\lambda_{hhh}$ and the measured parameters $m_h$ and $v$ receives a $10\%$ correction arising from top-loop~\cite{Hollik:2001px}. To isolate the NP contribution to $\delta\lambda_{hhh}$, this correction should be accounted for, taking into consideration $\delta\lambda_{hff}$ of the studied NP model:
\beqa
\delta\lambda_{hhh}=\frac{\lambda_{hhh}}{\lambda_{hhh}^{\rm SM}}-1-\frac{m_t^4\left(1+4\delta\lambda_{htt}\right)}{\pi^2v^2m_h^2}\,.
\eeqa
For brevity, we omit this correction in the following. The leading SM loop-correction to $\lambda_{hVV}$ is of order $1\%$~\cite{Kanemura:2004mg}: $\delta\lambda_{hVV}^{SM}=5m_t^2/\left(32\pi^2v^2\right)$. Additional loop-induced contributions to $\lambda_{hVV}$ (arising from, for example, Higgs-loop) are further suppressed and can be safely neglected. We also neglect all other NP loop-induced contributions to $\lambda_{hhh}$ which are found to be subdominant.

An analysis related to ours has recently been presented in Ref.~\cite{Gupta:2013zza}. The main focus in Ref.~\cite{Gupta:2013zza} is on a quantitative question: Within various NP scenarios, what is the maximal value for $\delta\lambda_{hhh}$ which is allowed by current experimental bounds on other Higgs couplings? The bound is then compared to future experimental prospects. In our study, the main focus is on a qualitative question: Assuming that deviations from the SM will be observed, how can one combine $\delta\lambda_{hhh}$ with other Higgs coupling measurements to support or exclude various relevant extensions of the SM? We further ask which specific features of these models will be probed. The two studies are complementary. Whenever the same models are considered, our results agree with the results obtained in Ref.~\cite{Gupta:2013zza}.

The plan of this paper is as follows. In Section~\ref{sec:pros} we review the experimental status and future prospects for measurements of $\lambda_{hhh}$ and $\lambda_{hVV}$. To demonstrate the power of the ratio $\delta\lambda_{hhh}/\delta\lambda_{hVV}$ we study it in several extensions of the SM scalar sector: the addition of a gauge singlet (Section~\ref{sec:singlet}), an extra $SU(2)_W$ doublet (Section~\ref{sec:doublet}) and additional $SU(2)_W$ triplets (Section~\ref{sec:triplets}). In Section~\ref{sec:EFT} we analyze the impact of general dimension six effective interaction for the SM $h$. We conclude in Section~\ref{sec:conc}.

\section{Experimental status and prospects}\label{sec:pros}
The first probe of the Higgs self coupling would come from Higgs pair-production, in which a virtual Higgs splits into two on-shell Higgs particles in the final state. (See, {\it e.g.}, Refs.~\cite{Dawson:1998py,Djouadi:1999ei,Muhlleitner:2000jj,Weiglein:2004hn,Djouadi:2005gi,Dolan:2012rv,Baglio:2012np}.) At the LHC, the cross section for $h$ pair production is roughly 1000 times smaller than the for single $h$. It is dominated by the gluon-gluon fusion (ggF) mechanism, and followed by the vector boson fusion (VBF) and Higgs-strahlung off vector-boson (VHH). The sensitivity of these channels to $\lambda_{hhh}$ has been extensively studied. (See, {\it e.g.}, Fig.~13 in Ref.~\cite{Baglio:2012np} and Ref.~\cite{Shao:2013bz}.) Whenever studying NP modifying the Higgs trilinear self coupling, one should bear in mind that two Higgs particles can also be produced without self-interactions. These background processes do not depend on $\lambda_{hhh}$, and interfere with the signal process. The ggF and VBF production channels exhibit destructive interference which suppresses the total $h$ pair production cross section for $\lambda_{hhh}^{\rm SM}\lesssim\lambda_{hhh}\lesssim3\lambda_{hhh}^{\rm SM}$.

As concerns the various decay channels, although the $4b$ final state is the dominant one, it suffers from huge QCD background. The most promising channel at the LHC is thought to be the rare decay $hh\rightarrow\gamma\gamma b\bar{b}$. The main backgrounds for this decay mode are QCD processes and a single Higgs production in association with top pair. Other possibly promising final states are $b\bar{b}\tau^-\tau^+$~\cite{Barr:2013tda,Dolan:2013rja,Maierhofer:2013sha} and $b\bar{b}W^+W^-$~\cite{Papaefstathiou:2012qe}. Different studies estimate the expected sensitivity for $\lambda_{hhh}$ at the LHC to be somewhere between 30\% and 50\% for $\sqrt{s}=14$~TeV and 3 ab${}^{-1}$ of integrated luminosity~\cite{ATLAS-collaboration:2012iza,Goertz:2013kp,Yao:2013ika,Barger:2013jfa}.

In $e^+e^-$ collisions the main production channels of Higgs pairs are double Higgs-strahlung off $Z$ bosons ($ZHH$, for $\sqrt{s}=500$~GeV) and double Higgs fusion ($\nu\nu HH$, for $\sqrt{s}\geq1$~TeV). (See, {\it e.g.}, Refs.~\cite{Weiglein:2004hn,Yasui:2002se,Castanier:2001sf,Djouadi:2007ik}.) To overcome the large background of the $4b$ final state high energy collisions are needed along with high $b$-tagging efficiency. Studies for a linear $e^+e^-$ collider find that a relative accuracy of $20~(10)\%$ is expected for CM collision energy of 500~GeV (1~TeV) and $\lag\simeq1{\rm ~ab}^{-1}$. However, Refs.~\cite{Dawson:2013bba,Tian:2013yda} quotes only $21\%$ for the expected accuracy at the ILC with $\sqrt{s}=1$~TeV and $\lag=1000$ fb${}^{-1}$. An ILC-based photon collider would give a poor sensitivity for $\lambda_{hhh}$ of only about $1\sigma$~\cite{Dawson:2013bba}. An $e^+e^-$ synchrotron at the di-Higgs threshold would presumably enable an accuracy of $28\%$ for $\lambda_{hhh}$~\cite{McCullough:2013rea}.

As for the Higgs couplings to heavy gauge bosons, the recent LHC Higgs measurements suggest that these are similar to the SM prediction. Yet, they are not tightly constrained. Using the most updated data given by the ATLAS and CMS collaborations~\cite{dataATLAS,dataCMS} we perform a naive minimum $\chi^2$ analysis and find the allowed range for $\lambda_{hVV}$, assuming custodial symmetry holds. Profiling over a universal coupling to fermions we find, within $95\%$ C.L.:
\beqa
-15\%&\lesssim&\delta\lambda_{hVV}\lesssim 5\%\,.
\eeqa
The expected sensitivity of future measurements to $\delta\lambda_{hVV}$ is given in Table~\ref{tab:expV} under the assumption of generation universal fermion couplings.
\begin{table}[h!]
\begin{center}
\begin{tabular}{|c|c|c|c|c|c|c|} \hline\hline
\rule{0pt}{1.2em}%
Experiment                                     & $\sqrt{s}$ [TeV]  & $\lag$ [fb${}^{-1}$]        & $\delta\lambda_{hVV}\lesssim$ \cr \hline\hline
LHC (ATLAS)~\cite{ATLAS-collaboration:2013}    & 14                 & 300                                 & $\left(2.5-3.3\right)\%$ \cr
LHC (CMS)~\cite{CMS_strategy}                  & 14                 & 300                                 & $\left(2.7-5.7\right)\%$ \cr
LHC (ATLAS)~\cite{ATLAS-collaboration:2013}    & 14                 & 3000                                & $\left(1.6-2.6\right)\%$ \cr
LHC (CMS)~\cite{CMS_strategy}                  & 14                 & 3000                                & $\left(1.0-4.5\right)\%$ \cr
ILC~\cite{Dawson:2013bba}                      & 0.25+0.5           & 250+500                             & $0.39\%$   \cr
ILC~\cite{Dawson:2013bba}                      & 0.25+0.5+1         & 250+500+1000                        & $0.21\%$   \cr
\hline\hline
\end{tabular}
\end{center}
\caption{Expected sensitivity to $\delta\lambda_{hVV}$ at the LHC and ILC. The upper and lower limits represent, respectively, a conservative and optimistic scenarios for the systematic errors at the LHC. A $0.5\%$ theoretical uncertainty is assumed for the expected results from the ILC. Generation universal couplings to fermions are assumed.}\label{tab:expV}
\end{table}

In the following we obtain the predictions of different NP models, giving special attention to the information that can be extracted from comparing $\delta\lambda_{hhh}$ and $\delta\lambda_{hVV}$. Whenever informative, we study also the Higgs coupling to fermion pairs.

\section{Doublet-singlet mixing}\label{sec:singlet}

The minimal extension of the SM Higgs sector is the addition of a single real gauge-singlet scalar $(\Phi_S)$ which mixes with the SM Higgs doublet $(\Phi_{\rm SM})$~\cite{Gupta:2013zza,Dolan:2012ac}. In the following we study the implications of such a mixing on the couplings of $h$, assuming a $Z_2$ symmetry under which $\Phi_S\rightarrow-\Phi_S$. We further elaborate on experimental constraints on the singlet-doublet mixing parameter space.

We consider the following scalar potential:
\beqa
{\cal V}&=&\mu^2\Phi_{\rm SM}^\dagger\Phi_{\rm SM}+\lambda\left(\Phi_{\rm SM}^\dagger\Phi_{\rm SM}\right)^2+m_S^2\Phi_S^2+\rho\Phi_S^4+\eta\Phi_{\rm SM}^\dagger\Phi_{\rm SM}\Phi_{\rm S}^2\,.
\eeqa
with
\beqa
\Phi_{\rm SM}=\begin{pmatrix}0 \\ \frac{1}{\sqrt{2}}\left(v+\phi_{\rm SM}\right)\end{pmatrix}\,,\;\;\;\;\;
\Phi_S=\frac{1}{\sqrt{2}}\left(\Lambda+\phi_S\right)\,.
\eeqa
In terms of the Lagrangian parameters, we have
\beqa
v^2=\frac{2m_S^2\eta-4\mu^2\rho}{4\lambda\rho-\eta^2}>0\,,\;\;\;\;\;\Lambda^2=\frac{2\eta\mu^2-4\lambda m_S^2}{4\lambda\rho-\eta^2}>0\,.
\eeqa
The physical spectrum contains a light $(h)$ and a heavy $(H)$ neutral scalar, which are linear combinations of the SM and singlet fields:
\beq\label{eq:rotation}
h=c_\alpha\phi_{\rm SM}+s_\alpha\phi_S\,,\;\;\;\;\;H=-s_\alpha\phi_{\rm SM}+c_\alpha\phi_S\,,
\eeq
where $c_\alpha\equiv\cos\alpha$ and $s_\alpha\equiv\sin\alpha$. The masses and the mixing angle are given by:
\beqa\label{eq:physparameters}
m_{H,h}^2=\lambda v^2+\rho \Lambda^2\pm\sqrt{\left(\rho \Lambda^2-\lambda v^2\right)^2+\eta^2v^2\Lambda^2}\,,\;\;\;\;\;\tan\alpha=\frac{\eta v\Lambda}{m_H^2-2\lambda v^2}\,.
\eeqa
If $H$ is very heavy, the light scalar $h$ has SM-like properties. In this limit $v\ll\Lambda$ and
\beqa
m_h^2\simeq\left(2\lambda-\frac{\eta^2}{2\rho}\right)v^2\,,\;\;\;\;
m_H^2\simeq2\rho \Lambda^2\gg m_h^2\,,\;\;\;\;
s_\alpha\simeq\frac{\eta}{2\rho}\frac{v}{\Lambda}\ll1\,.
\eeqa
%

\subsection{Results}
The couplings of $h$ to the weak gauge bosons and to the charged fermions are different from the SM predictions due to the small mixing with the singlet:
\beqa
\frac{\lambda_{hVV}}{\lambda_{hVV}^{\rm SM}}&=&\frac{\lambda_{hff}}{\lambda_{hff}^{\rm SM}}=c_\alpha\simeq\left(1-\frac{1}{2}s^2_\alpha\right)\,.
\eeqa
The trilinear self coupling of the light scalar is
\beq\label{eq:singletlambda}
\frac{\lambda_{hhh}}{\lambda_{hhh}^{\rm SM}}=\left(c_\alpha^3-s_\alpha^3\frac{v}{\Lambda}\right)\simeq\left(1-\frac{3}{2}s^2_\alpha\right)\,.
\eeq
This gives, to leading order in $s_\alpha$,
\beq
\frac{\delta\lambda_{hhh}}{\delta \lambda_{hVV}}\simeq3\,.
\eeq
We thus learn the following points on the Higgs couplings in the decoupling limit:
\begin{itemize}
\item The couplings of $h$ to pairs of weak gauge bosons and of charged fermions deviate from the SM predictions. The deviation is small, of order $s_\alpha^2$, and negative.
\item The deviation is the same for  the weak bosons and for the charged fermions, $\delta\lambda_{hVV}=\delta\lambda_{hff}$.
\item The $h$ trilinear self coupling deviates from the SM. The deviation is small, of order $s_\alpha^2$, and negative.
\item $\delta\lambda_{hhh}/\delta \lambda_{hVV}=3$ at leading order. This relation is independent of the NP parameters, and provides a decisive test for this extension of the SM Higgs sector.
\item In principle, the singlet VEV $\Lambda$ can be extracted by combining the information from $\lambda_{hhh}$ and $\lambda_{hVV}$. In practice, the required accuracy is beyond reach.
\end{itemize}
The predictions of the SM with additional scalar singlet are given in the doublet-singlet row of Table~\ref{tab:couplings}. The experimental constraints on $s_\alpha$ are described in the following section. These imply that, for the doublet-singlet mixing, a deviation of $-0.11\lesssim\delta\lambda\leq0$ is allowed within $95\%$ C.L..

\subsection{Experimental constraints}
Recent Higgs measurements at the LHC constrain the doublet-singlet mixing angle. Using $\delta\lambda_{hVV}=\delta\lambda_{hff}$, we find the Best Fit Point (BFP), $s_\alpha\simeq0.28$, and the upper bound, $s_\alpha\lesssim0.51$ at $95\%$ C.L.. The one-dimensional compatibility of the SM prediction with the best-fit value is $40\%$.

The doublet-singlet mixing and the presence of an extra heavy scalar change the prediction for the oblique ElectroWeak (EW) parameters with respect to their SM values~\cite{Peskin:1991sw}:
\beqa
\delta X=s_\alpha^2\left[X_S(m_H)-X_S(m_h)\right]\,,
\eeqa
with $X_S$ the scalar loop contribution to the parameter $X=S,T$, as given in Appendix~C of Ref.~\cite{Hagiwara:1994pw}. To find the resulting bounds we use the combined EW fit for the $S$ and $T$ parameters (setting $U=0$) from Ref.~\cite{Baak:2012kk}:
\beqa
S=0.05\pm0.09\,,\;\;\;T=0.08\pm0.07\,,\;\;\;(\rho_{\rm corr}=0.91)\,.
\eeqa
For $m_H=1$~TeV, electroweak precision measurements (EWPM) then imply
\beq
s_\alpha\lesssim0.27\ {\rm at}\ 95\%\ {\rm C.L.}\,.
\eeq
This bound is compatible with the limit of detectability at the LHC with $\sqrt{s}=14$ TeV and 100 fb${}^{-1}$ of integrated luminosity, as obtained in Ref.~\cite{Gupta:2013zza}, and with the current Higgs data from the LHC.

\section{Doublet-doublet mixing}\label{sec:doublet}
We consider the CP conserving Two Higgs Doublet Model (2HDM). We use the notations of Ref.~\cite{Gunion:2002zf}. The scalar potential for two $SU(2)_W$ doublet scalar fields, $\Phi_{1,2}$ is given by
\beqa
\mathcal{V}&=&m_{11}^2\Phi_1^\dagger\Phi_1+m_{22}^2\Phi_2^\dagger\Phi_2-\left[m_{12}^2\Phi_1^\dagger\Phi_2+{\rm ~h.c.}\right]\nonumber\\
& &+\frac{1}{2}\lambda_1\left(\Phi_1^\dagger\Phi_1\right)^2+\frac{1}{2}\lambda_2\left(\Phi_2^\dagger\Phi_2\right)^2
+\lambda_3\left(\Phi_1^\dagger\Phi_1\right)\left(\Phi_2^\dagger\Phi_2\right)+\lambda_4\left(\Phi_1^\dagger\Phi_2\right)\left(\Phi_2^\dagger\Phi_1\right)\nonumber\\
& &+\left[\frac{1}{2}\lambda_5\left(\Phi_1^\dagger\Phi_2\right)^2+\lambda_6\left(\Phi_1^\dagger\Phi_1\right)\left(\Phi_1^\dagger\Phi_2\right)
+\lambda_7\left(\Phi_2^\dagger\Phi_2\right)\left(\Phi_1^\dagger\Phi_2\right)+{\rm ~h.c.}\right]\,.
\eeqa
We denote the VEVs of the two doublets by $v_{1,2}$, with $v^2=v_2^2+v_2^2$ and $\tan\beta\equiv v_2/v_1$. The spectrum of the Higgs sector contains two CP-even neutral states $(h,H)$, a CP-odd neutral state $(A)$ and a charged scalar $(H^\pm)$. The angle $\alpha$ is the rotation angle from the real neutral components of the two doublets to the physical CP-even mass eigenstates. If $H$ is very heavy, the light scalar $h$ has SM-like properties. In this limit the remaining scalars, $H,A$ and $H^\pm$, are mass degenerate up to corrections of $\mathcal{O}\left(v^2/m_A^2\right)$. We represent their common mass scale by the mass of the pseudoscalar, $m_A$. In the decoupling limit~\cite{Gunion:2002zf}
\beq
\cos\left(\beta-\alpha\right)\simeq\frac{\hat{\lambda}v^2}{m_A^2}\ll1\,,
\eeq
where
\beqa
\hat{\lambda}&\equiv&\frac{1}{2}s_{2\beta}\left[\lambda_1c^2_\beta-\lambda_2s^2_\beta-\left(\lambda_3+\lambda_4+\lambda_5\right)c_{2\beta}\right]-\lambda_6c_\beta c_{3\beta}-\lambda_7s_\beta s_{3\beta}\,.
\eeqa
%

\subsection{Results}
The coupling of the light scalar $h$ to the weak gauge bosons is different from the SM prediction due to the misalignment between the Higgs basis (defined by the angle $\beta$) and the mass basis (defined by the angle $\alpha$):
\beq
\frac{\lambda_{hVV}}{\lambda_{hVV}^{\rm SM}}=\sin\left(\beta-\alpha\right)
\simeq1-\frac{1}{2}\cos^2\left(\beta-\alpha\right)\,.
\eeq
Its trilinear self coupling is~\cite{Gupta:2013zza}
\beq\label{eq:singletlambda}
\frac{\lambda_{hhh}}{\lambda_{hhh}^{\rm SM}}
\simeq1-\frac{2m_A^2}{m_h^2}\cos^2\left(\beta-\alpha\right)\,.
\eeq
The couplings of $h$ to fermion pairs are model dependent. In the type II 2HDM they read
\beqa\label{eq:fermions}
\frac{\lambda_{htt}}{\lambda_{htt}^{\rm SM}}&=&\frac{\cos\alpha}{\sin\beta}
=\sin\left(\beta-\alpha\right)+\frac{\cos\left(\beta-\alpha\right)}{\tan\beta}\,,\nonumber\\
\frac{\lambda_{hbb}}{\lambda_{hbb}^{\rm SM}}&=&-\frac{\sin\alpha}{\cos\beta}
=\sin\left(\beta-\alpha\right)-\tan\beta\cos\left(\beta-\alpha\right)\,.
\eeqa
We obtain
\beq
\frac{\delta\lambda_{hhh}}{\delta\lambda_{hVV}}\simeq\frac{4m_A^2}{m_h^2}\,.
\eeq

We thus learn the following points:
\begin{itemize}
\item The couplings of $h$ to $ZZ$ and $WW$ deviate from the SM prediction. The deviation is small, of order $\cos^2\left(\beta-\alpha\right)$, and negative.
\item The Higgs couplings to fermions are model dependent. In the type II 2HDM, for moderate values of $\tan\beta$, the deviation from the SM prediction is of order $\cos\left(\beta-\alpha\right)$.
\item The $h$ trilinear self coupling deviates from the SM prediction. The deviation is of order \\
$\left(m_A^2/m_h^2\right)\cos^2\left(\beta-\alpha\right)\simeq\cos\left(\beta-\alpha\right)$, and negative.
\item At leading order $\delta\lambda_{hhh}/\delta\lambda_{hVV}=4m_A^2/m_h^2\gg1$. Thus, the deviation in the trilinear coupling is much larger than the deviation in the coupling to the weak gauge bosons.
\item Using the measured value of $\delta\lambda_{hhh}/\delta\lambda_{hVV}$,
    one will be able to cleanly extract the value of $m_A$, the mass scale of the second Higgs doublet which, if indeed the decoupling limit applies, will be out of direct reach in the LHC experiments.
\end{itemize}
The predictions of the 2HDM are given in the doublet-doublet row of Table~\ref{tab:couplings}. Taking $\hat\lambda\simeq1$ and $m_A\simeq500$~GeV, we find $\delta h_{VV}\sim-0.03$, while $\delta\lambda_{hhh}\simeq-1.9$. This will enhance the di-Higgs production at the LHC (with $\sqrt{s}=14$~TeV) by a factor of five. For $\hat\lambda\simeq1$ and $m_A\simeq1$ TeV, we find $\delta\lambda_{hVV}\simeq-0.001$, while $\delta\lambda_{hhh}\simeq-0.46$.

We note that, at tree level, the results for the MSSM are the same as those of type II 2HDM. The two-loop $\mathcal{O}\left(\alpha_s\alpha_t\right)$ corrections to $\lambda_{hhh}$ within the MSSM are calculated in Ref.~\cite{Brucherseifer:2013qva}.

\subsection{Experimental constraints}
We study the experimental constraint on the Type I and Type II 2HDM using the latest LHC Higgs data. We consider two independent parameters: $-1\leq\sin\left(\beta-\alpha\right)\leq1$ and $0.5\leq\tan\beta\leq65$, verifying that $\abs{\alpha}\leq\pi/2$ is maintained. We assume that the heavy scalars do not affect the various measurements. Our results are given in Table~\ref{tab:exp2HDM}.

\begin{table}[h!]
\begin{center}
\begin{tabular}{|c|c|c|} \hline\hline
\rule{0pt}{1.2em}                                     &Type I                                       & Type II                        \cr \hline\hline
BFP for $[\frac{\lambda_{hVV}}{\lambda_{hVV}^{\rm SM}},\frac{\lambda_{huu}}{\lambda_{huu}^{\rm SM}},\frac{\lambda_{hdd}}{\lambda_{hdd}^{\rm SM}}]$ &$[0.96,0.98,0.98]$                           & $[0.94,1.00,-1.03]$                         \cr
SM compatibility                                      &69\%                                            & 72\%                                           \cr
$\delta\lambda_{hVV}$ at $95\%$ C.L                   &$-15\%\lesssim\delta\lambda_{hVV}\lesssim 0$        &$-23\%\lesssim\delta\lambda_{hVV}\lesssim 0$\cr
\hline\hline
\end{tabular}
\end{center}
\caption{The implications of the LHC Higgs data on the Type I and Type II 2HDMs. The allowed range for $\delta\lambda_{hVV}$ is obtained by scanning over $0.5\leq\tan\beta\leq65$.}\label{tab:exp2HDM}
\end{table}
A more detailed analysis for the experimentally allowed parameter space in various 2HDM types can be found in, {\it e.g.,} Refs.~\cite{Ferreira:2013qua,Eberhardt:2013uba,Celis:2013ixa,Chang:2013ona,Ilisie:2013cxa,Belanger:2013xza,Cheung:2013rva}.

\section{Doublet-triplets mixing}\label{sec:triplets}
Additional scalar fields in the $SU(2)_W$ triplet representation can account for neutrino masses via the Type~III See-saw mechanism~(for a review, see \cite{GonzalezGarcia:2002dz}). When adding a single $SU(2)_W$ triplet with $Y=1$, custodial symmetry is violated at tree level. EWPM establish $\rho=m_W^2/(m_Z^2\cos^2\theta_W)=1$ to a very good accuracy, leading to severe constraints on the mixing with such a triplet:
\beqa
\sin^2\alpha\leq5\times10^{-3}\,.
\eeqa
Since $\delta\lambda_{hhh}\sim\delta\lambda_{hVV}\sim\sin^2\alpha$, all the deviations in the couplings are, at most, at the few permil level.

To avoid this tight constraint we study the Georgi-Machacek model~\cite{Georgi:1985nv,Chanowitz:1985ug}, in which the Higgs sector contains the SM Higgs doublet and three real triplets with hypercharges $Y=-1,0,1$. We follow the notations of Ref.~\cite{Gunion:1989ci}. The three triplets are combined into one matrix field
\beqa
\chi=\begin{pmatrix}\chi^0&\xi^+&\chi^{++}\\ \chi^-&\xi^0&\chi^+\\ \chi^{--}&\xi^-&\chi^{0*}\end{pmatrix}\,,
\eeqa
with $\avg{\chi}={\rm diag}\left(b,b,b\right)$. This particular choice of VEVs guarantees that $\rho=1$ at tree level. The SM doublet VEV is denoted as $\avg{\phi}=a/\sqrt{2}$.
We find
\beqa\label{eq:thetaH}
v^2=a^2+8b^2\,,
\eeqa
leaving the ratio of the two VEVs a free parameter. It is useful to define an angle $\theta_H$ such that
\beqa
c_H\equiv\cos\theta_H\equiv a/v\,,\;\;\;
s_H\equiv\sin\theta_H\equiv\sqrt{8}b/v\,.
\eeqa

Altogether, there are ten physical real scalars. Under custodial $SU(2)$, they decompose into a five-plet, a triplet and two singlets. The latter two are
\beqa\label{eq:cussin}
H_1^0={\rm Re}\left[\phi^0\right]\,,\;\;\;\;H_1^{0'}=\frac{1}{\sqrt{3}}\left(\chi^0+\chi^{0*}+\xi^0\right)\,.
\eeqa
Assuming that the Higgs potential obeys the custodial symmetry, the two states of Eq. (\ref{eq:cussin}) mix (only) among themselves, forming two mass eigenstates $h,H$. We define the rotation angle to the mass eigenstates, $\alpha$, via
\beq
h=c_\alpha H_1^0+s_\alpha H_1^{0'}\,,\;\;\;\;\;H=-s_\alpha H_1^0+c_\alpha H_1^{0'}\,,
\eeq
with $c_\alpha\equiv\cos\alpha$, $s_\alpha\equiv\sin\alpha$.

As concerns the Yukawa couplings, only the SM doublet couples to charged fermion pairs at tree level, with $Y_{H_1^0ff}=Y_{f}^{\rm SM}/c_H$. The Yukawa couplings of the mass eigenstate scalars are then given by
\beq\label{eq:hcoup}
\frac{\lambda_{hff}}{\lambda_{hff}^{\rm SM}}=\frac{c_\alpha}{c_H}\,,\;\;\;\frac{\lambda_{Hff}}{\lambda_{hff}^{\rm SM}}=-\frac{s_\alpha}{c_H}\,.
\eeq
The couplings to pairs of weak bosons are deduced from the gauge kinetic terms in the Lagrangian~\cite{Belanger:2013xza,Gunion:1989ci}:
\beq\label{eq:hHvv}
\frac{\lambda_{hVV}}{\lambda_{hVV}^{\rm SM}}=\left(c_\alpha c_H+\frac{2\sqrt{2}}{\sqrt{3}}s_\alpha s_H\right)\,,\;\;\;\frac{\lambda_{HVV}}{\lambda_{hVV}^{\rm SM}}=\left(-s_\alpha c_H+\frac{2\sqrt{2}}{\sqrt{3}}c_\alpha s_H\right)\,.
\eeq
It is then clear that the SM limit is reached when both $s_\alpha\ll1$ and $s_H\ll1$, in which case $h$ has SM-like properties.

To obtain the scalar masses and self interactions one needs to specify the scalar potential. Imposing the custodial symmetry and a $Z_2$ symmetry under which $\chi\rightarrow-\chi$, the most general scalar potential takes the form:
\beqa
{\cal V}&=&\lambda_1\left(2\phi^\dagger\phi-a^2\right)^2+\lambda_2\left({\rm Tr}\left[\chi^\dagger\chi\right]-3b^2\right)^2
+\lambda_3\left(2\phi^\dagger\phi-a^2+{\rm Tr}\left[\chi^\dagger\chi\right]-3b^2\right)^2\no\\
&+&\lambda_4\left(2\phi^\dagger\phi{\rm Tr}\left[\chi^\dagger\chi\right]-2{\rm Tr}\left[\Phi^\dagger\tau_i\Phi\tau_j\right]{\rm Tr}\left[\chi^\dagger t_i\chi t_j\right]\right)^2
+\lambda_5\left(3{\rm Tr}\left[\chi^\dagger\chi\chi^\dagger\chi\right]-\left({\rm Tr}\left[\chi^\dagger\chi\right]\right)^2\right)^2\,,
\eeqa
with $\Phi=\left(\sigma_1\phi^*,\phi\right)$. Here $\tau_a/2$ are the $2\times 2$ representation matrices of $SU(2)_W$ and $t_a$ are the $3\times 3$ representation matrices of $SU(2)_W$ in cartesian coordinates. Since the different custodial multiplets do not mix with each other, all the members of the five-plet have well defined mass, and likewise all the members of the triplet. Reaching the perturbativity limit $\lambda_4\simeq4\pi$ these can be as heavy as $m_3\simeq870$~GeV and $m_5\simeq1.5$~TeV. The singlet mass matrix is:
\beqa
\mathbb{M}^2_{H_1^0,H_1^{0'}}=\begin{pmatrix}8c_H^2\lambda_{13}&2\sqrt{6}c_Hs_H\lambda_3\\2\sqrt{6}c_Hs_H\lambda_3&3s_H^2\lambda_{23}\end{pmatrix}v^2\,,
\eeqa
where $\lambda_{13}\equiv\lambda_1+\lambda_3$, $\lambda_{23}\equiv\lambda_2+\lambda_3$. Stability of the potential requires $\lambda_{13}>0$ and $\lambda_{23}>0$, while $\lambda_{13}\lambda_{23}>\lambda_3^2$ should hold for $m_{1,2}^2>0$. The masses and the mixing angle are given by
\beqa
m_{h,H}^2&=&\frac{v^2}{2}\left(8c_H^2\lambda_{13}+3\lambda_{23}s_H^2\pm\sqrt{\left(8c_H^2\lambda_{13}-3s_H^2\lambda_{23}\right)^2+96c_H^2s_H^2\lambda_3^2}\right)\,,\no\\
\tan\alpha&=&\frac{2\sqrt{6}c_Hs_H\lambda_{3}}{m_{h}^2/v^2-8c_H^2\lambda_{13}}\,.
\eeqa
Clearly, in the limit of $\lambda_3=0$ no mixing occurs. If, in addition, $s_H=0$ a $U(1)$ symmetry is restored under which $\chi$ can be rotated by a pure phase. In this case, $m_H=0$. Yet, we find that $m_{H}\simeq380$~GeV is allowed for $\lambda_2\simeq4\pi$. We elaborate more on this issue in Sec.~\ref{subsec:GMEXP}.
The trilinear scalar couplings are given by
\beqa
\frac{\lambda_{hhh}}{6v}&=&4c_\alpha^3c_H\lambda_{13}+\sqrt{6}c_\alpha^2s_\alpha s_H\lambda_3+4c_\alpha s_\alpha^2c_H\lambda_3+\sqrt{6}s_\alpha^3s_H\lambda_{23}\,,\no\\
\frac{\lambda_{HHH}}{6v}&=&\sqrt{6}c_\alpha^3s_H\lambda_{23}-4c_\alpha^2s_\alpha c_H\lambda_3+\sqrt{6}c_\alpha s_\alpha^2s_H\lambda_3-4s_\alpha^3c_H\lambda_{13}\,.
\eeqa
%

\subsection{Results}
In this subsection we obtain simplified results for $\delta\lambda_{hhh},\,\delta\lambda_{hVV}$ and $\delta\lambda_{hff}$, considering two different limits: First, the limit of small mixing, $s_\alpha^2\ll s_H^2$, and, second, the limit of small triplet VEV, $s_H^2\ll s_\alpha^2$.
It is convenient to define
\beq
\Delta^2\equiv m_{H}^2-m_{h}^2,\ \ \ \tilde{\Delta}^2\equiv\left(3s_H^2\lambda_{23}-8c_H^2\lambda_{13}\right)v^2,
\eeq
and to use
\beq\label{eq:lamthr}
\lambda_3=-\frac{s_\alpha\Delta^2}{2\sqrt{6}c_Hs_Hv^2}\,.
\eeq
%

\subsubsection{Small mixing}
We consider the case of $s_\alpha^2\ll1$. For $s_\alpha=0$, the LHC Higgs data allow $0.88<c_H<1$ within $95\%$ C.L., where $c_H=0.97$ is the BFP (with SM compatibility of 46\%). We thus take
\beq\label{eq:trihiea}
0<s_\alpha^2\ll s_H^2\ll1\,.
\eeq
Using (\ref{eq:lamthr}) we find:
\beqa
m_{h}^2/v^2&\simeq&8c_H^2\lambda_{13}-
\frac{2s_\alpha^2\Delta^4}{\tilde{\Delta}^2v^2}\,,\no\\
\lambda_{hhh}&\simeq&24c_H\lambda_{13}v-\frac{3s_\alpha^2
\left(\Delta^2+12c_H^2v^2\lambda_{13}\right)}{c_Hv}\,.
\eeqa
These relations give:
\beqa
\frac{\lambda_{hhh}}{\lambda_{hhh}^{\rm SM}}&\simeq&\frac{1}{c_H}\left[1-s_\alpha^2\left(
\frac{3}{2}+\frac{1}{8}\frac{\Delta^2}{c_H^2\lambda_{13}v^2}
\left(1-\frac{2\Delta^2}{\tilde{\Delta}^2}\right)\right)\right]\no\\
&\simeq&\frac{1}{c_H}\left[1-s_\alpha^2
\left(\frac{5}{2}-\frac{m_H^2}{m_h^2}\right)\right]\,,
\eeqa
where we use the fact that, to leading order, $\Delta^2\simeq\tilde{\Delta}^2$.

As concerns the $h$ couplings to the charged fermions and to the weak gauge bosons, Eqs.~\eqref{eq:hcoup} and \eqref{eq:hHvv} give
\beqa
\frac{\lambda_{hff}}{\lambda_{hff}^{\rm SM}}&\simeq&\frac{1}{c_H}\left(1-\frac{1}{2}s_\alpha^2\right)\,,\no\\
\frac{\lambda_{hVV}}{\lambda_{hVV}^{\rm SM}}&\simeq&c_H\left(1-\frac{1}{2}s_\alpha^2+\frac{2\sqrt{2}}{\sqrt{3}}t_Hs_\alpha\right)\,.
\eeqa
%

We learn the following points, which apply for the hierarchy (\ref{eq:trihiea}):
\begin{itemize}
\item The couplings of $h$ to charged fermion pairs and to weak gauge bosons deviate from their SM prediction with $\lambda_{hVV}\leq\lambda_{hVV}^{\rm SM}$ but $\lambda_{hff}\geq\lambda_{hff}^{\rm SM}$.
\item The trilinear self coupling deviates from its SM prediction. To leading order, we find:
\beq
\delta\lambda_{hhh}\simeq\delta\lambda_{hff}\simeq-\delta\lambda_{hVV}
\simeq\frac12 s_H^2\,.
\eeq
\item $\lambda_{hhh}$ contains information on the mass of the other singlet scalar, $H$. In principle, by combining the experimental information on $\delta\lambda_{hhh},\delta\lambda_{hVV}$ and $\delta\lambda_{hff}$, $m_H$ can be deduced.
\end{itemize}
In Sec.~\ref{subsec:GMEXP} we further elaborate on current experimental constraints on the $s_\alpha-s_H$ parameter space, from both $h$ related measurements and null searches for the other scalar, $H$.

The predictions of the Georgi-Machacek model in the limit of small mixing are given in the "doublet-triplet with $\alpha\ll1$" row of Table~\ref{tab:couplings}. In this regime, $0\leq\delta\lambda_{hhh}\lesssim13\%$ is allowed within 95\% C.L. by current experimental bounds.

\subsubsection{Small triplet VEV}
We consider the case of $s_H^2\ll1$. For $s_H=0$, the LHC data allow $s_\alpha<0.51$ within 95\% C.L., and $s_\alpha=0.28$ is the BFP (with SM compatibility of 40\%). We thus take
\beq\label{eq:hiesh}
0<s_H^2\ll s_\alpha^2\ll1\,.
\eeq
We find:
\beqa
m_{h}^2&\simeq&8\lambda_{13}v^2\,,\no\\
\lambda_{hhh}&\simeq&24v\lambda_{13}
\left(c_\alpha^3+s_\alpha^2c_\alpha\frac{\lambda_3}{\lambda_{13}}\right)\,.
\eeqa
Using (\ref{eq:lamthr}) we obtain
\beqa
\frac{\lambda_3}{\lambda_{13}}&\simeq&-\frac{4s_\alpha\Delta^2}{\sqrt{6}c_Hm_h^2},
\eeqa
which gives
\beqa
\lambda_{hhh}&\simeq&24v\lambda_{13}\left(c_\alpha^3-s_\alpha^3c_\alpha
\frac{4}{\sqrt{6}}\frac{\Delta^2}{m_h^2}\right)\,.
\eeqa
We thus find that
\beqa
\frac{\lambda_{hhh}}{\lambda_{hhh}^{\rm SM}}\simeq c_\alpha^3\left(1-\frac{4t_\alpha^3c_\alpha}{\sqrt{6}}\frac{\Delta^2}{m_h^2}\right)\,.
\eeqa

As concerns the $h$ couplings to the charged fermions and to the weak gauge bosons, Eqs.~\eqref{eq:hcoup} and \eqref{eq:hHvv} give
\beqa
\frac{\lambda_{hff}}{\lambda_{hff}^{\rm SM}}&\simeq&c_\alpha\left(1+\frac{1}{2}s_H^2\right)\,,\no\\
\frac{\lambda_{hVV}}{\lambda_{hVV}^{\rm SM}}&\simeq&c_\alpha\left(1-\frac{1}{2}s_H^2+\frac{2\sqrt{2}}{\sqrt{3}}t_\alpha s_H\right)\,,\no\\
\eeqa

In the limit of small $s_H$ the second scalar has to be light since its mass is proportional to $s_H^2$:
\beqa
m_{H}^2&\simeq&3s_H^2v^2\left(\lambda_{23}-\frac{\lambda_3^2}{\lambda_{13}}\right)\,.
\eeqa
Yet, if also $s_\alpha$ is small enough, its couplings to both gauge bosons and fermions are very small and it might escape detection. The complete experimental analysis is done in the next subsection.

We learn the following points, which apply for the hierarchy (\ref{eq:hiesh}):
\begin{itemize}
\item The couplings of $h$ to charged fermion pairs and to the weak gauge bosons deviate from their SM prediction, with
    \beq
    \delta\lambda_{hVV}\simeq\delta\lambda_{hff}<0\,.
    \eeq
\item The $h$ trilinear self coupling deviates from its SM prediction, with
\beq
\frac{\delta\lambda_{hhh}}{\delta\lambda_{hVV}}\simeq3\,,
\eeq
\item $\lambda_{hhh}$ contains information on the mass of the other singlet scalar, $H$. In principle, by combining $\delta\lambda_{hhh},\delta\lambda_{hVV}$ and $\delta\lambda_{hff}$, $m_H$ can be deduced.
\end{itemize}
The predictions of the Georgi-Machacek model in the limit of small $s_H$ are given in the "doublet-triplet with $\theta_H\ll1$" row of Table~\ref{tab:couplings}. In this regime, $-39\%\lesssim\delta\lambda_{hhh}\leq0$ is allowed within 95\% C.L. by current experimental bounds.

\subsection{Experimental constraints}\label{subsec:GMEXP}

\begin{figure}[h!]\label{fig:GMLHC}
  \begin{center}
      \subfigure[$m_H=100$~GeV]{\scalebox{0.9}{\includegraphics[width=2.5in]{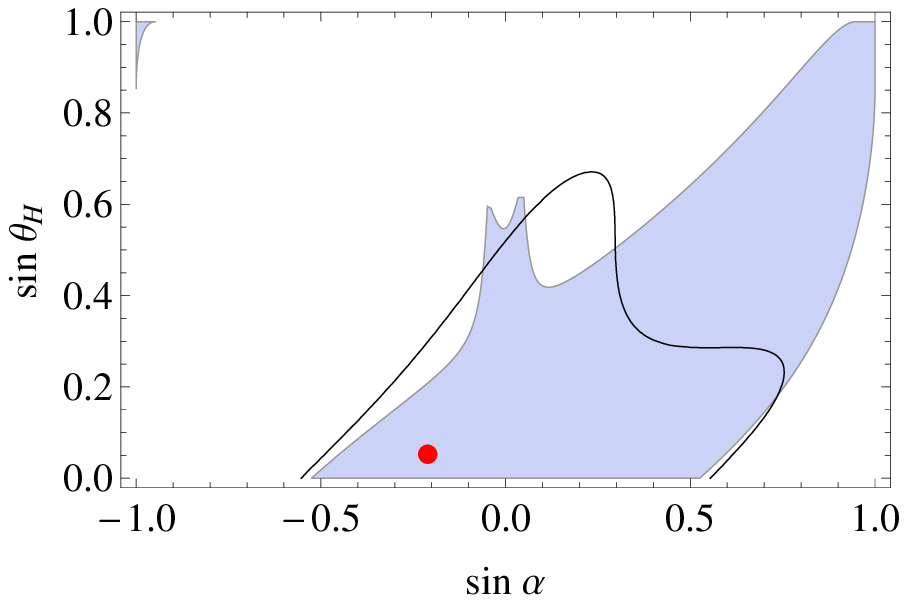}}\label{fig:GM100}}
      \subfigure[$m_H=240$~GeV]{\scalebox{0.9}{\includegraphics[width=2.5in]{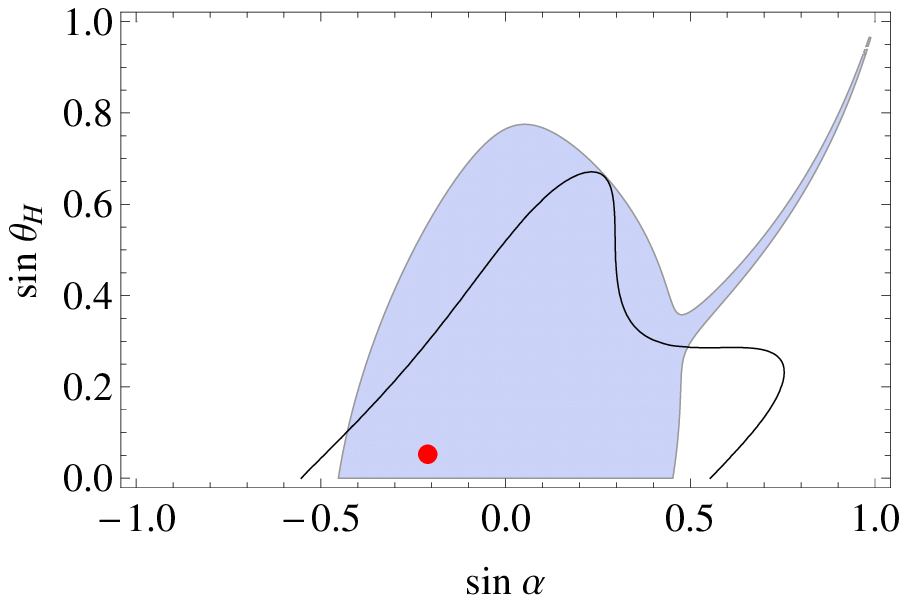}}\label{fig:GM240}}
      \caption{The allowed 95\% C.L. parameter space of the Georgi-Machacek model. The LHC Higgs data allow the region inside the black curve. The direct $H$ searches from LEP (left, for $M_H=100$~GeV) and LHC (right, for $M_H=240$~GeV) allow the blue region. The red circle represents the Higgs data Best Fit Point.}
  \end{center}
\end{figure}

Assuming $h$ is the recently-discovered Higgs boson, LHC Higgs data prefers both $s_\alpha\sim1$ and $s_H\sim1$. The BFP for the Higgs measurements is $\left(s_H,s_\alpha\right)=\left(0.05,-0.21\right)$ with $69\%$ compatibility for the SM. Yet, $s_H$ as large as 0.62, and $s_\alpha\simeq 0.70$ are allowed at $95\%$ C.L. (scanning over the other parameters.) Fig.~\ref{fig:GMLHC} shows the allowed region in the $s_H-s_\alpha$ plane from the LHC Higgs data.

As concerns $H$, since its coupling to gauge bosons is not suppressed with $s_\alpha\times s_H$, it could be produced directly at the LEP collider using the $e^+e^-\rightarrow Z^*\rightarrow ZH$ process for $m_H\lesssim115$~GeV (as shown in~\cite{Belanger:2013xza}). The relevant decay channels in this case are the $b\bar{b}$ and $VV^*$ final states. We note that in some of the relevant parameter space the decay modes of $H$ do not resemble those of the SM Higgs, since its coupling to gauge bosons can be much larger than its coupling to $b$ quarks. We consider $m_H=100$~GeV (avoiding $h\rightarrow HH$ decays) and use
\beqa\label{eq:LEP}
\frac{\sigma_{\rm prod}\Gamma_{H\rightarrow X}\Gamma_{\rm tot}^{\rm SM}}{\sigma_{\rm prod}^{\rm SM}\Gamma_{H\rightarrow X}^{\rm SM}\Gamma_{\rm tot}}\leq\mu^{\rm LEP}_X
\eeqa
where $X=WW^*,ZZ^*,b\bar{b}$ and $\mu^{\rm LEP}_{X}$ is the corresponding LEP bound (normalized to the SM Branching ratio), taken from Refs.~\cite{Schael:2006cr,Achard:2003jb}. We further use
\beqa
\frac{\sigma_{\rm prod}}{\sigma_{\rm prod}^{\rm SM}}&=&\left(\frac{\lambda_{HVV}}{\lambda_{hVV}^{\rm SM}}\right)^2\,,\no\\
\frac{\Gamma_{H\rightarrow X}}{\Gamma_{H\rightarrow X}^{\rm SM}}&=&\left(\frac{\lambda_{HX}}{\lambda_{hX}^{\rm SM}}\right)^2\,.
\eeqa
The combined LEP and LHC Higgs constraints for $m_H=100$~GeV are shown in Fig.~\ref{fig:GM100}. We note that the $VV^*$ decay mode affects only a small portion of the parameter space around $s_\alpha\sim0$ and $s_H\sim1$.

In the mass region $m_{h}\leq m_{H}\leq2m_{h}$, direct searches at the LHC constrain the BR($H\to WW)$ and BR($H\to ZZ)$. In this case, we use
\beqa
\frac{\left(\frac{\lambda_{Hff}}{\lambda_{hff}^{\rm SM}}\right)^2\sigma_{ggF}^{\rm SM}+\left(\frac{\lambda_{HVV}}{\lambda_{hVV}^{\rm SM}}\right)^2\sigma_{VBF}^{\rm SM}}{\sigma_{ggF}^{\rm SM}+\sigma_{VBF}^{\rm SM}}\times\left(\frac{\lambda_{HVV}}{\lambda_{hVV}^{\rm SM}}\right)^2\frac{\Gamma_{h,{\rm tot}}^{\rm SM}}{\Gamma_{H,{\rm tot}}}\leq\mu^{\rm LHC}_{VV}\,,
\eeqa
where $\mu^{\rm LHC}_{VV}$ is the corresponding LHC bound taken from~\cite{Chatrchyan:2013yoa}, and the SM production cross sections and branching ratios are taken from~\cite{Heinemeyer:2013tqa}. In this case, we consider $m_H=240$~GeV. Our results are shown in Fig.~\ref{fig:GM240}. If $H$ is heavier than $2m_{h}$ no bounds exist since its dominant decay mode is to light Higgs boson pairs. We comment that TeVatron data yields no further constraints on the relevant parameter space.

\section{Higher dimensional effective interaction}\label{sec:EFT}
Heavy NP can induce an effective $\left(\Phi_{\rm SM}^\dagger\Phi_{\rm SM}\right)^3$ interaction, which may lead to a first-order EW phase transition~\cite{Grojean:2004xa}. (See also Ref.~\cite{Bonnet:2011yx,Bonnet:2012nm}.) If the heavy states do not mix with the SM Higgs, the Higgs couplings to gauge bosons and fermions do not deviate from their SM values. This happens, for example, when a heavy singlet field that does not acquire a VEV is added to the SM Higgs sector. In this section we study the influence of this nonrenormalizable interaction. We consider the following Higgs potential:
\beq
{\cal V}=\mu^2\left(\Phi_{\rm SM}^\dagger\Phi_{\rm SM}\right)+\lambda\left(\Phi_{\rm SM}^\dagger\Phi_{\rm SM}\right)^2+\frac{\rho}{\Lambda^2}\left(\Phi_{\rm SM}^\dagger\Phi_{\rm SM}\right)^3\,,
\eeq
and assume $\mu\ll\Lambda$. The potential has a minimum at
\beq
v^2=\frac{2\Lambda^2\lambda}{3\rho}\left(-1+\sqrt{1-\frac{3\rho\mu^2}{\lambda^2\Lambda^2}}\right)\simeq-\frac{\mu^2}{\lambda}\left(1+\frac{3\rho}{4\lambda^2}\frac{\mu^2}{\Lambda^2}\right)\,,
\eeq
with $\rho>0$. We further find:
\beqa
m_h^2&=&2v^2\lambda+3\rho\frac{v^4}{\Lambda^2}\,,\nonumber\\
\lambda_{hhh}&=&3\frac{m_h^2}{v}+6\rho\frac{v^3}{\Lambda^2}\,.\nonumber\\
\eeqa
%

\subsection{Results}
It might be that the only low energy imprint of heavy NP is the dimension six Higgs interaction. If this is the case, we learn the following:
\begin{itemize}
\item The $h$ couplings to the weak gauge bosons and to the charged fermions are the same as predicted by the SM. NP loop-corrections to these couplings, arising from $\delta\lambda_{hhh}$, are expected to be below the percent level.
\item The trilinear coupling deviates from the SM prediction:
\beqa
\delta\lambda_{hhh}=\frac{2\rho v^4}{m_h^2\Lambda^2}>0\,.
\eeqa
\item A measurement of $\lambda_{hhh}$ is the only way to reveal information on the NP that modifies the Higgs potential. The size of $\delta\lambda_{hhh}$ would allow an estimated upper bound on the scale of the new physics.
\end{itemize}

The predictions of the SM plus effective dimension six interaction are given in the $\left(\phi^\dagger\phi\right)^3$ row of Table~\ref{tab:couplings}. Additional class of models that modify only the Higgs trilinear self coupling can be found in Ref.~\cite{Abel:2013mya}.

\section{Discussion and conclusions}\label{sec:conc}

\begin{table}[h!]
\begin{center}
\begin{tabular}{|c|c|c|c|} \hline\hline
\rule{0pt}{1.2em}%
Model                                         & $\delta \lambda_{hVV}$              & $\delta \lambda_{hhh}$                   & $\delta\lambda_{hhh}/\delta \lambda_{hVV}$ \cr \hline\hline
Doublet-singlet mixing                        & $-s_\alpha^2/2$                     & $-3s_\alpha^2/2$                         & $3$                                  \cr
Doublet-doublet mixing                        & $-\hat{\lambda}^2v^4/(2m_A^4)$      & $-2\hat{\lambda}^2v^4/(m_h^2m_A^2)$      & $4m_A^2/m_h^2$                       \cr
Doublet-triplets mixing with $s_\alpha\ll1$   & $-s_H^2/2$                          & $s_H^2/2$                                & $-1$                                 \cr
Doublet-triplets mixing with $s_H\ll1$        & $-s_\alpha^2/2$                     & $-3s_\alpha^2/2$                         & $3$                                  \cr
$(\phi^\dagger\phi)^3$                        & $0$                                 & $2\rho v^4/(\Lambda^2m_h^2)$             & $\infty$                             \cr
\hline\hline
\end{tabular}
\end{center}
\caption{Predictions of various extensions of the SM scalar sector for the $hVV$ and $hhh$ couplings.}\label{tab:couplings}
\end{table}

Within the SM, the trilinear self-coupling of the Higgs boson fulfills $\lambda_{hhh}^{\rm SM}=3m_h^2/v$. A future measurement of this coupling will test this relation. In case that it is violated, the measurement will constitute a significant probe of the mechanism that breaks the EW symmetry. Current measurements of the Higgs couplings to gauge bosons and to charged fermions do not show any significant deviation from the SM predictions. Current direct searches for new particles do not give any evidence that such particles exist within the direct reach of the LHC. It is thus possible that a measurement of the trilinear coupling will give a first hint for NP that modifies the Higgs potential.

In this work, we study the decoupling limit of several well motivated extensions of the Higgs sector. We provide general expressions for the resulting Higgs couplings and show that their deviations from the SM predictions exhibit well-defined patterns related to the structure of the Higgs potential. Our results are summarized in Table \ref{tab:couplings}. We found that the ratio $\delta\lambda_{hhh}/\delta\lambda_{hVV}$ is often larger than one, namely the deviation in the trilinear coupling is more significant. Moreover, the ratio is often independent of the details of the new physics, and consequently provides information that is uniquely clean:
\begin{itemize}
\item Singlet-doublet mixing or triplet-doublet mixing in the small $s_H$ limit of the Georgi-Machacek model predict $\delta\lambda_{hhh}\simeq3\delta\lambda_{hVV}$ accompanied with $\delta\lambda_{hVV}\simeq\delta\lambda_{hff}$.
\item The small $s_\alpha$ limit of the Georgi-Machacek model predicts $\delta\lambda_{hhh}\simeq\delta\lambda_{hff}\simeq-\delta\lambda_{hVV}$.
\item Two Higgs doublet models predict $\delta\lambda_{hhh}\gg\delta\lambda_{hVV}$. In this case, the mass of the other scalars can be determined via $m_A^2\simeq(\delta_{\lambda_{hhh}}/\delta\lambda_{hVV})(m_h^2/4)$.
\item If a deviation is observed in $\lambda_{hhh}$ but neither in the $h$ coupling to the weak gauge boson, nor in its couplings to the charged fermions, then it could be that the only significant effect of the NP is to generate the dimension six term $(\Phi^\dagger\Phi)^3$.
\end{itemize}

The Higgs couplings to $W^+W^-$, $ZZ$, $\tau^+\tau^-$ and, perhaps, $b\bar b$ and $\mu^+\mu^-$ will be measured with better and better accuracy in the coming years. If deviations from the SM are observed, then a measurement of the Higgs trilinear self coupling, which is still far from reach, will become highly desired. We demonstrated the power of combining the $hVV$ and $hf\bar f$ coupling measurements with a measurement of the $hhh$ coupling in distinguishing between various models of NP and shedding light on the scale of NP.

\vspace{1cm}
\begin{center}
{\bf Acknowledgements}
\end{center}
We thank Avital Dery and Yotam Soreq for many fruitful discussions. YN is the Amos de-Shalit chair of theoretical physics. YN is supported by the Israel Science Foundation, by the I-CORE program of the Planning and Budgeting Committee and the Israel Science Foundation (grant number 1937/12), and by the German-Israeli foundation for scientific research and development (GIF).


\end{document}